\documentclass{article}

\usepackage[eandd, final]{neurips_2026}

\usepackage[utf8]{inputenc} 
\usepackage[T1]{fontenc}    
\usepackage{hyperref}       
\usepackage{url}            
\usepackage{graphicx}       
\usepackage{booktabs}       
\usepackage{amsmath}        
\usepackage{amsfonts}       
\usepackage{amssymb}        
\usepackage{nicefrac}       
\usepackage{microtype}      
\usepackage{xcolor}         
\usepackage{subcaption}     
\usepackage{multirow}       
\usepackage{tabularx}       
\usepackage{flafter}        
\usepackage{listings}       

\hypersetup{
  pdftitle={SWE-GPU-Bench: Can Language Models Solve Real-World GPU Software Engineering Tasks?},
  pdfauthor={Feng Chen, Binbin Liu, Wenhan Han, Yin Zheng},
  colorlinks=true,
  linkcolor=blue!50!black,
  citecolor=blue!50!black,
  urlcolor=blue!50!black,
}

\lstdefinestyle{promptstyle}{
  basicstyle=\ttfamily\scriptsize,
  breaklines=true,
  breakatwhitespace=false,
  columns=fullflexible,
  keepspaces=true,
  frame=single,
  framerule=0.25pt,
  rulecolor=\color{black!35},
  backgroundcolor=\color{black!2},
  xleftmargin=2pt,
  xrightmargin=2pt,
  aboveskip=0.6\baselineskip,
  belowskip=0.6\baselineskip
}

\lstdefinestyle{cmdstyle}{
  basicstyle=\ttfamily\scriptsize,
  breaklines=true,
  columns=fullflexible,
  keepspaces=true,
  frame=single,
  framerule=0.25pt,
  rulecolor=\color{black!35},
  backgroundcolor=\color{black!2},
  xleftmargin=2pt,
  xrightmargin=2pt,
  moredelim=**[is][\color{orange!75!black}\bfseries]{@H}{H@}
}

\definecolor{gold accent}{RGB}{15,118,110}
\definecolor{gold bg}{RGB}{240,253,250}
\definecolor{agent accent}{RGB}{194,65,12}
\definecolor{agent bg}{RGB}{255,247,237}

\lstdefinestyle{goldcode}{
  basicstyle=\ttfamily\scriptsize,
  breaklines=true,
  columns=fullflexible,
  keepspaces=true,
  frame=single,
  framerule=0.5pt,
  rulecolor=\color{gold accent!70},
  backgroundcolor=\color{gold bg!55},
  xleftmargin=3pt,
  xrightmargin=3pt,
  aboveskip=2pt,
  belowskip=0pt,
  moredelim=**[is][\color{gold accent}\bfseries]{@G}{G@}
}

\lstdefinestyle{agentcode}{
  basicstyle=\ttfamily\scriptsize,
  breaklines=true,
  columns=fullflexible,
  keepspaces=true,
  frame=single,
  framerule=0.5pt,
  rulecolor=\color{agent accent!70},
  backgroundcolor=\color{agent bg!55},
  xleftmargin=3pt,
  xrightmargin=3pt,
  aboveskip=2pt,
  belowskip=0pt,
  moredelim=**[is][\color{agent accent}\bfseries]{@A}{A@}
}

\title{SWE-GPU-Bench: Can Language Models Solve Real-World GPU Software Engineering Tasks?}

\author{%
  Feng Chen\quad
  Binbin Liu\quad
  Wenhan Han\quad
  Yin Zheng\thanks{Corresponding author. Email: yzheng3xg@gmail.com}\\
}

\begin{document}

\maketitle

\begin{abstract}
Existing benchmarks for GPU programming with language models evaluate whether a model can generate an isolated kernel, operator, or standalone CUDA program. Most GPU software, however, is developed inside existing repositories, where a single change often spans CUDA kernels, C++ host code, Python bindings, build files, and tests. We introduce SWE-GPU-Bench, a benchmark of 417 tasks drawn from merged pull requests in 30 GPU repositories, including CUTLASS, CCCL, FlashInfer, Triton, and TensorFlow. Working in a pre-configured environment, a model receives a problem statement and the repository at the base commit to generate a patch, which we evaluate using the pull request's test commands. The tasks cover bug fixes, feature implementations, and performance optimizations, with a small subset of performance tasks (Perf-Replay) also including benchmark commands that measure the speedup of patches that pass the correctness tests. Over a quarter of the tasks edit more than four files, and over half edit files in two or more languages. We evaluate frontier open-weight and proprietary models in both the oracle and agent settings. The strongest evaluated configurations, GPT-6-Astra and GPT-6.1-Sol in Codex CLI, resolve 58.3\% and 58.0\% of tasks, respectively.
\end{abstract}

\section{Introduction}

Recent benchmarks evaluate language models on GPU code generation. KernelBench and TritonBench, for example, test whether a model can generate a functionally correct and performant kernel or operator from a specification~\citep{kernelbench,tritonbench}. In practice, however, GPU software engineering takes place inside large repositories, where updating a kernel often requires coordinated changes across host launch logic, language bindings, build configurations, and tests.

\begin{figure}[!htb]
  \centering
  \includegraphics[width=\textwidth]{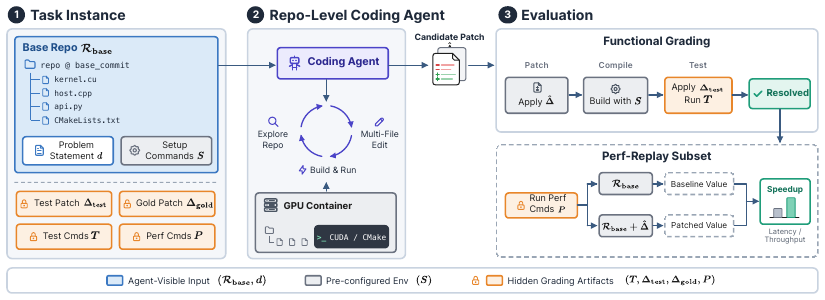}
  \caption{Overview of SWE-GPU-Bench evaluation. \textcircled{\scriptsize 1}~Each task instance is a tuple $(\mathcal{R}_{\mathrm{base}}, d, S, T, P, \Delta_{\mathrm{test}}, \Delta_{\mathrm{gold}})$. After the environment is prepared with $S$, the model or agent receives only the repository $\mathcal{R}_{\mathrm{base}}$ at the base commit and the problem statement $d$, while $T$, $\Delta_{\mathrm{test}}$, $\Delta_{\mathrm{gold}}$, and $P$ remain hidden. \textcircled{\scriptsize 2}~A coding agent explores the repository, edits files, and builds and runs code in a GPU container to output a candidate patch $\hat{\Delta}$. \textcircled{\scriptsize 3}~Functional grading applies $\hat{\Delta}$ and $\Delta_{\mathrm{test}}$ and runs $T$; for Perf-Replay tasks, patches passing $T$ are benchmarked with $P$ against $\mathcal{R}_{\mathrm{base}}$ to measure speedup.}
  \label{fig:overview}
\end{figure}

Repository-level benchmarks such as SWE-bench capture software maintenance for Python projects~\citep{swebench}, but GPU repositories introduce two further challenges. First, their codebases span multiple languages and build systems, with CUDA and C++ sources compiled through CMake, Bazel, or setuptools and invoked from Python. Second, executing tests requires matching the target CUDA toolkit, architecture-specific compiler flags, and GPU hardware. Addressing these cross-language build dependencies and hardware execution requirements is central to real-world GPU engineering, yet existing repository-level GPU benchmarks remain limited to small-scale code translation~\citep{parevalrepo} or narrow inference-engine tuning~\citep{isobench}, rather than general software maintenance across bug fixes, feature implementations, and performance optimizations.

We introduce SWE-GPU-Bench (Figure~\ref{fig:overview}), constructing tasks from merged pull requests (PRs) across 30 GPU repositories.\footnote{Data: \url{https://huggingface.co/datasets/Froze/SWE-GPU-Bench}. Code: \url{https://github.com/chenfengshijie/SWE-GPU-Bench}. We report the current release of 417 tasks (378 general-set and 39 Perf-Replay tasks); the benchmark dataset is actively maintained and will be continuously updated.} For each task, we first run task-specific setup commands to prepare the build environment, then provide the model with the repository at the base commit and a problem statement distilled from the PR, and finally apply the PR's test changes and run its test commands to grade the generated patch. To construct the benchmark, we filter candidate PRs with deterministic rules and an LLM judge, use a coding agent to synthesize setup and test commands, and retain only tasks whose tests fail before applying the gold patch and pass afterward, yielding 417 tasks. For a subset of performance tasks (Perf-Replay), we also synthesize benchmark commands to measure the speedup of patches that pass functional tests.

We evaluate frontier open-weight and proprietary models under two settings following SWE-bench and SWE-Perf~\citep{swebench,sweperf}: an \emph{oracle} setting, where a model receives the files modified by the gold patch and generates a patch in a single pass, and an \emph{agent} setting, where a coding agent iteratively explores the repository, edits code, and invokes the build and test toolchain. Across evaluated models, iterative agent interaction yields substantially higher resolution rates than single-pass oracle generation even when the oracle is given the exact target files. Whereas earlier models frequently fail to compile across heterogeneous GPU build systems, the latest frontier models achieve nearly 100\% compilation accuracy—reflecting rapid progress on repository-level GPU builds—yet the strongest configurations, GPT-6-Astra and GPT-6.1-Sol in Codex CLI, resolve 58.3\% and 58.0\% of tasks, with remaining failures concentrated in test execution.

Our contributions are:
\begin{itemize}
  \item \textbf{A realistic repository-level GPU benchmark.} We introduce SWE-GPU-Bench, spanning bug fixes, feature implementations, and performance optimizations across production GPU libraries, compilers, and frameworks. Its tasks require coordinated edits across CUDA kernels, C++ host logic, Python bindings, and heterogeneous build systems, complemented by the Perf-Replay subset for correctness-gated speedup evaluation.
  \item \textbf{An automated, execution-verified construction pipeline.} To handle heterogeneous GPU build environments, our five-stage pipeline combines rule- and LLM-based filtering with an interactive coding agent that synthesizes task-specific setup, test, and benchmark commands inside GPU containers. Every task is verified via strict fail-to-pass replay in a clean environment and audited by human reviewers against solution leakage.
  \item \textbf{A multi-stage empirical evaluation and failure analysis.} Evaluating frontier open-weight and proprietary models across oracle and agent settings, we decompose performance into patch application, compilation, test execution, and runtime speedup. Our analysis demonstrates the necessity of execution feedback over static file localization, quantifies accuracy degradation across multi-file and multi-language edits, and reveals a sharp gap between functional correctness and end-to-end GPU optimization.
\end{itemize}

\section{SWE-GPU-Bench}
\label{sec:benchmark}

SWE-GPU-Bench is a repository-level benchmark of 417 GPU software maintenance tasks constructed from merged PRs, encompassing a general set of 378 tasks for functional evaluation and a Perf-Replay subset of 39 tasks for runtime speedup evaluation (Section~\ref{sec:perf-replay}).

\subsection{Task Formulation}
\label{sec:formulation}

Each task is packaged as $(\mathcal{R}_{\mathrm{base}}, d, S, T, P, \Delta_{\mathrm{test}}, \Delta_{\mathrm{gold}})$. We first check out the repository $\mathcal{R}_{\mathrm{base}}$ at the base commit of the PR, remove all subsequent git history, and run the setup commands $S$ to prepare the build environment. Providing a pre-configured environment decouples repository-level code maintenance from container provisioning. The model or agent is then given $\mathcal{R}_{\mathrm{base}}$ and the problem statement $d$ and outputs a candidate patch $\hat{\Delta}$. The grading artifacts remain hidden from the model: the test patch $\Delta_{\mathrm{test}}$ (tests added or modified by the PR), the gold patch $\Delta_{\mathrm{gold}}$ (all remaining changes), the test commands $T$, and, for Perf-Replay tasks, the benchmark commands $P$.

To evaluate $\hat{\Delta}$, we apply $\hat{\Delta}$ to $\mathcal{R}_{\mathrm{base}}$, revert any candidate modifications to files targeted by $\Delta_{\mathrm{test}}$, apply $\Delta_{\mathrm{test}}$, and run $T$ to compile and execute the PR's tests. A task is marked \emph{resolved} if all commands in $T$ exit with code~0, consistent with the fail-to-pass verification during construction (Section~\ref{sec:construction}). For Perf-Replay tasks that pass $T$, we additionally run $P$ to evaluate runtime speedup over $\mathcal{R}_{\mathrm{base}}$ (Section~\ref{sec:perf-replay}). Over all attempted tasks, we report three metrics: \textbf{Patch} (producing a non-empty, cleanly applicable patch), \textbf{Compile} (building the patched repository without compilation errors), and \textbf{Test} (passing $T$ to resolve the task).

\subsection{Benchmark Construction}
\label{sec:construction}

\begin{figure}[t]
  \centering
  \includegraphics[width=\textwidth]{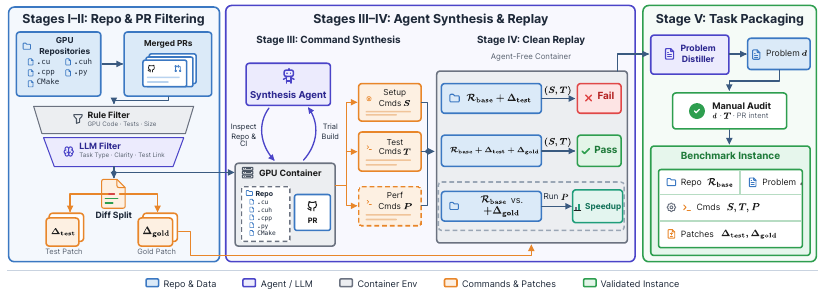}
  \caption{Construction pipeline of SWE-GPU-Bench. \textbf{Stages I--II (repository and PR filtering):} merged PRs from GPU repositories pass a rule filter (default-branch merge, linked issue, GPU code, tests, bounded code size) and an LLM filter (task type, clarity, test linkage), and each PR diff is split into the test patch $\Delta_{\mathrm{test}}$ and the gold patch $\Delta_{\mathrm{gold}}$. \textbf{Stage III (command synthesis):} a synthesis agent in a GPU container alternates between inspecting the repository and CI and running trial builds to produce setup commands $S$, test commands $T$, and, for performance PRs, benchmark commands $P$. \textbf{Stage IV (clean replay):} in an agent-free container, $(S, T)$ must fail on $\mathcal{R}_{\mathrm{base}}{+}\Delta_{\mathrm{test}}$ and pass once $\Delta_{\mathrm{gold}}$ is also applied; for performance PRs, $P$ is additionally run on $\mathcal{R}_{\mathrm{base}}$ and $\mathcal{R}_{\mathrm{base}}{+}\Delta_{\mathrm{gold}}$ to verify the reference speedup. \textbf{Stage V (task packaging):} an LLM distills the problem statement $d$, a manual audit verifies $d$, $T$, and the PR intent, and the result is packaged as $(\mathcal{R}_{\mathrm{base}}, d, S, T, P, \Delta_{\mathrm{test}}, \Delta_{\mathrm{gold}})$. Dashed elements apply only to performance PRs.}
  \label{fig:construction}
\end{figure}

Because many PRs in GPU repositories lack targeted tests, bundle unrelated changes, or depend on custom CI environments, we construct tasks through a five-stage pipeline (Figure~\ref{fig:construction}); Appendix~\ref{app:benchmark-details} provides full filtering criteria and benchmark statistics.

\paragraph{Stage I: Repository selection and PR collection.} We select actively maintained GPU repositories from official organizations and core maintainers, spanning CUDA libraries, GPU compilers such as Triton and TVM, ML frameworks, and inference engines. Across these repositories, we collect merged PRs along with their metadata, commit histories, and diffs.

\paragraph{Stage II: Filtering.} Deterministic rule filters retain PRs that merge into the default branch, reference a resolved issue, contain no binary diffs, and modify both tests and at least one CUDA, C++, or GPU build file. To focus on self-contained maintenance tasks with isolatable fail-to-pass tests, we also bound the number of modified files and lines, excluding broad refactors and release merges (Appendix~\ref{app:filtering}). An LLM judge (GPT-5.2) then classifies each remaining PR (bug fix, feature, or performance) and scores its GPU relevance, problem clarity, test linkage, and hardware reproducibility on NVIDIA A100 or H100 GPUs. Candidates rated accepted or borderline advance to synthesis; the LLM judge serves strictly as a candidate filter, whereas final task admission requires executable verification (Stage~IV) and manual audit (Stage~V). Each candidate's diff is partitioned into the test patch $\Delta_{\mathrm{test}}$ (test files and test-support targets) and the gold patch $\Delta_{\mathrm{gold}}$ (all remaining changes).

\paragraph{Stage III: Command synthesis.} For each candidate PR, a synthesis agent runs inside an A100 or H100 GPU container matched to the target architecture (e.g., H100 for Hopper-specific features such as \texttt{sm\_90a}). The agent alternates between inspecting repository metadata, build scripts, and CI workflows and executing trial builds and test runs, iteratively refining its commands until it synthesizes setup commands $S$, test commands $T$ targeting the PR's tests, and, for performance PRs, benchmark commands $P$. Command synthesis is the most compute-intensive stage of construction (prompt in Appendix~\ref{app:prompt-commands}; cost details in Appendix~\ref{app:cost}).

\paragraph{Stage IV: Clean replay.} We replay $S$ and $T$ in a clean, agent-free GPU container. A candidate is admitted only if $T$ fails on $\mathcal{R}_{\mathrm{base}}$ with $\Delta_{\mathrm{test}}$ applied and passes with exit code~0 once $\Delta_{\mathrm{gold}}$ is also applied. For performance PRs, we additionally execute $P$ on $\mathcal{R}_{\mathrm{base}}$ and $\mathcal{R}_{\mathrm{base}}+\Delta_{\mathrm{gold}}$ to confirm that the reference patch yields a measurable improvement (Section~\ref{sec:perf-replay}).

\paragraph{Stage V: Task packaging.} An LLM distills the problem statement $d$ from the PR title, description, and diff under constraints that prohibit naming modified files, internal symbols, or edit steps unless required by public test interfaces (Appendix~\ref{app:prompt-statement}). Human reviewers then audit every candidate passing clean replay to ensure that $d$ leaks no solution details, that $T$ exercises the intended change, and that $d$, $T$, and the PR intent are mutually consistent (Appendix~\ref{app:human-validation}). Accepted candidates are packaged as $(\mathcal{R}_{\mathrm{base}}, d, S, T, P, \Delta_{\mathrm{test}}, \Delta_{\mathrm{gold}})$ (format in Appendix~\ref{app:format}), with $P$ included for Perf-Replay tasks.

\subsection{Features of SWE-GPU-Bench}
\label{sec:features}

\begin{figure}[t]
  \centering
  \begin{minipage}[c]{0.46\textwidth}
    \centering
    \includegraphics[width=\linewidth]{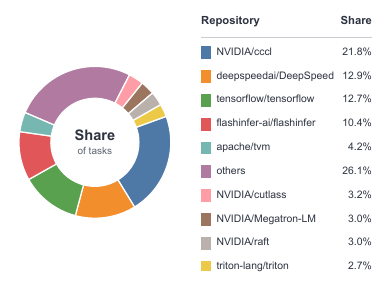}
    \caption{Distribution of tasks by repository, with the smallest shares grouped as \emph{others}.}
    \label{fig:repos}
  \end{minipage}\hfill
  \begin{minipage}[c]{0.51\textwidth}
    \centering
    \captionof{table}{Mean and maximum of task attributes over all tasks. Appendix~\ref{app:stats} reports percentiles.}
    \label{tab:stats}
    \small
    \begin{tabular}{@{}llrr@{}}
      \toprule
      & & Mean & Max \\
      \midrule
      Problem statement & \# Words & 236.6 & 2{,}636 \\
      \midrule
      \multirow{3}{*}{Gold patch} & \# Files & 3.7 & 62 \\
                                  & \# Functions & 6.0 & 166 \\
                                  & \# Lines & 178.6 & 3{,}227 \\
      \midrule
      \multirow{2}{*}{Test patch} & \# Files & 1.8 & 13 \\
                                  & \# Lines & 126.0 & 927 \\
      \midrule
      Test commands & \# Lines & 6.8 & 102 \\
      \bottomrule
    \end{tabular}
  \end{minipage}
\end{figure}

\begin{figure}[t]
  \centering
  \begin{subfigure}[t]{0.47\textwidth}
    \centering
    \includegraphics[width=\linewidth]{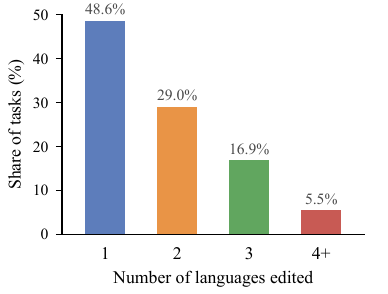}
    \caption{Languages per gold patch.}
    \label{fig:lang-bar}
  \end{subfigure}\hfill
  \begin{subfigure}[t]{0.47\textwidth}
    \centering
    \includegraphics[width=\linewidth]{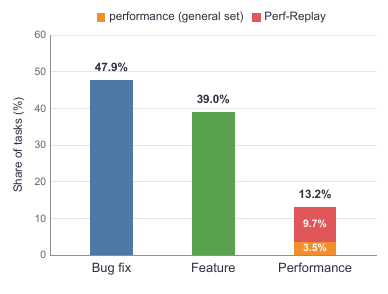}
    \caption{Task types.}
    \label{fig:task-bar}
  \end{subfigure}
  \caption{Composition of SWE-GPU-Bench, in percent of tasks. (a) Number of languages edited by the gold patch. (b) Distribution of task types, with the performance bar split into general-set and Perf-Replay tasks.}
  \label{fig:composition}
\end{figure}

\paragraph{Production GPU repositories.} Tasks are drawn from 30 repositories (Figure~\ref{fig:repos}) spanning four layers of the GPU software stack: CUDA libraries (CCCL, CUTLASS, RAFT), compilers and DSLs (Triton, TVM, XLA, TileLang), ML frameworks (TensorFlow, JAX, DeepSpeed, Megatron-LM), and inference engines (FlashInfer, LMDeploy, llama.cpp). Models must therefore localize edits within large-scale, production-grade repositories.

\paragraph{Multi-file, cross-language edits.} Gold patches frequently require coordinated changes across files and language boundaries: over a quarter of the tasks modify more than four files, and more than half span two or more languages—typically coupling low-level CUDA or C++ device code with host bindings in Python and build configurations in CMake or Bazel (Figure~\ref{fig:lang-bar}).

\paragraph{Diverse maintenance categories.} Following the primary classification from Stage~II, the benchmark spans bug fixes, feature implementations, and performance optimizations (Figure~\ref{fig:task-bar}), with hardware-compatibility updates, build-infrastructure changes, and refactorings grouped under features.

\paragraph{Architecture-aware executable verification.} Every task packages executable setup and test commands that compile the affected targets and run the PR's tests inside a GPU container (Table~\ref{tab:stats}). Typical workflows configure CMake, Bazel, or custom build scripts for a specific compute capability (such as \texttt{sm\_80} or \texttt{sm\_90a}) and invoke targeted GoogleTest or pytest suites, as illustrated by the CUTLASS case study in Appendix~\ref{app:case-study}.

\subsection{Perf-Replay}
\label{sec:perf-replay}

For performance-optimization PRs, unit tests verify functional correctness rather than runtime speedup. For the subset of performance PRs that include a reproducible benchmark workload, we construct Perf-Replay by synthesizing benchmark commands $P$ configured to output a scalar value for the target performance metric (such as latency, throughput, or kernel execution time). Performance tasks without a standalone benchmark remain in the general set and are evaluated through their correctness tests $T$.

Evaluating a candidate patch on Perf-Replay applies a correctness gate: the patch must first pass all functional tests in $T$. For patches that pass $T$, we execute $P$ on both $\mathcal{R}_{\mathrm{base}}$ and the patched repository to obtain their scalar measurements and compute the relative improvement ratio over $\mathcal{R}_{\mathrm{base}}$. Because standardized performance evaluation across heterogeneous GPU workloads and metrics remains an open challenge, we report task-level relative changes to support correctness-gated performance analysis.

\section{Experimental Setup}
\label{sec:setup}

\paragraph{Oracle setting.} Following the oracle retrieval setting of SWE-bench and the oracle setting of SWE-Perf~\citep{swebench,sweperf}, we provide the model with the problem statement, the repository README, and the full contents of the non-test files modified by the gold patch. The model outputs search-and-replace edits in a single pass, which are applied via exact block matching and converted into a unified diff (prompt in Appendix~\ref{app:prompt-oracle}). Providing the target files isolates edit generation from repository-wide localization, offering a controlled reference setting in which the edit locations are known.

\paragraph{Agent setting.} We run the GPT models in Codex CLI and all other models in OpenCode (prompt in Appendix~\ref{app:prompt-agent}). Operating in the pre-configured container environment (Section~\ref{sec:formulation}), each agent is given a single run per task to autonomously inspect, edit, build, and test the repository, after which we record its final \texttt{git diff} as the candidate patch $\hat{\Delta}$.

\paragraph{Models and evaluation protocol.} Table~\ref{tab:main-results} lists the evaluated models, and Appendix~\ref{app:models} provides their model identifiers, harness configurations, and evaluation coverage across dataset revisions. Models in both the oracle and agent settings are evaluated on a single NVIDIA A100 or H100 GPU matched to each task's target architecture under a single pass@1 run per configuration.

\section{Results}
\label{sec:results}

\subsection{Main Evaluation Results}

Table~\ref{tab:main-results} summarizes the main evaluation results across the three pipeline stages (\textbf{Patch}, \textbf{Compile}, and \textbf{Test}). Among agent configurations, GPT-6-Astra and GPT-6.1-Sol in Codex CLI achieve the highest resolution rates (\textbf{Test}) at 58.3\% and 58.0\%, followed by GPT-5.6-Sol in Codex CLI at 56.7\% and DeepSeek-V4.1-Flash in OpenCode at 55.3\%. In the oracle setting, GPT-6-Astra achieves the highest resolution rate at 49.2\%, followed by GPT-6.1-Sol at 47.6\% and GPT-5.6-Sol at 38.5\%.

\begin{table}[!htb]
  \centering
  \small
  \caption{Main results on SWE-GPU-Bench. \textbf{Patch}: producing a non-empty, applicable patch; \textbf{Compile}: building the patched repository without compilation errors; \textbf{Test}: passing the test commands $T$ to resolve the task. Best value per setting in bold.}
  \label{tab:main-results}
  \begin{tabular}{@{}llrrr@{}}
    \toprule
    Setting & Model & Patch (\%) & Compile (\%) & Test (\%) \\
    \midrule
    \multirow{16}{*}{Oracle}
      & GPT-6-Astra         & \textbf{100.0} & \textbf{100.0} & \textbf{49.2} \\
      & GPT-6.1-Sol         & \textbf{100.0} & \textbf{100.0} & 47.6 \\
      & GPT-5.6-Sol         & \textbf{100.0} & \textbf{100.0} & 38.5 \\
      & GPT-5.2             & 79.2 & 48.3 & 26.0 \\
      & DeepSeek-V4.1-Flash & \textbf{100.0} & \textbf{100.0} & 24.3 \\
      & DeepSeek-V4-Pro     & \textbf{100.0} & \textbf{100.0} & 21.4 \\
      & Claude Sonnet 4     & 75.2 & 49.6 & 20.8 \\
      & GPT-5-Codex         & 77.1 & 49.2 & 20.4 \\
      & GLM-5               & 60.3 & 36.8 & 17.2 \\
      & Kimi K2.5           & 68.4 & 42.1 & 14.4 \\
      & GPT-5.1-Codex       & 52.2 & 31.6 & 12.9 \\
      & DeepSeek-V3.1       & 51.7 & 36.4 & 11.0 \\
      & DeepSeek-V4-Flash   & 75.6 & 68.9 & 8.6 \\
      & DeepSeek-R1         & 54.1 & 38.8 & 8.6 \\
      & DeepSeek-V3.2       & 41.6 & 25.8 & 8.1 \\
      & Gemini 2.5 Pro      & 19.1 & 10.5 & 4.8 \\
    \midrule
    \multirow{7}{*}{Agent (Codex)}
      & GPT-6-Astra         & \textbf{100.0} & \textbf{100.0} & \textbf{58.3} \\
      & GPT-6.1-Sol         & \textbf{100.0} & \textbf{100.0} & 58.0 \\
      & GPT-5.6-Sol         & 98.0 & 98.0 & 56.7 \\
      & GPT-5.5             & \textbf{100.0} & 95.2 & 48.9 \\
      & GPT-5-Codex         & 85.0 & 47.0 & 33.0 \\
      & GPT-5.1-Codex       & 74.7 & 57.8 & 33.0 \\
      & GPT-5 mini          & 87.8 & 56.7 & 30.0 \\
    \cmidrule(l){2-5}
    \multirow{6}{*}{Agent (OpenCode)}
      & DeepSeek-V4.1-Flash & \textbf{100.0} & \textbf{100.0} & 55.3 \\
      & DeepSeek-V4-Pro     & 94.0 & 94.0 & 43.3 \\
      & DeepSeek-V3.1       & 85.1 & 58.5 & 29.0 \\
      & Kimi K2.5           & 85.0 & 53.0 & 27.0 \\
      & DeepSeek-V3.2       & 86.7 & 56.1 & 26.5 \\
      & DeepSeek-V4-Flash   & 88.2 & 88.2 & 26.3 \\
    \bottomrule
  \end{tabular}
\end{table}

\paragraph{Models achieve higher resolution rates in the agent setting.} Every model evaluated in both settings achieves a substantially higher resolution rate when run as an interactive agent than under single-pass oracle generation, even though the oracle is given the exact files modified by the gold patch. Specifically, GPT-6-Astra increases from 49.2\% to 58.3\%, GPT-6.1-Sol from 47.6\% to 58.0\%, GPT-5.6-Sol from 38.5\% to 56.7\%, GPT-5.1-Codex from 12.9\% to 33.0\%, and DeepSeek-V3.2 from 8.1\% to 26.5\%. The gain from iterative execution feedback is especially pronounced for the DeepSeek-V4 models: in single-pass oracle generation without compiler or test feedback, DeepSeek-V4.1-Flash (24.3\%) and DeepSeek-V4-Pro (21.4\%) trail GPT-6-Astra (49.2\%) by more than 24 percentage points, whereas in the agent setting DeepSeek-V4.1-Flash more than doubles its resolution rate to 55.3\% (and DeepSeek-V4-Pro to 43.3\%), narrowing the gap with the leading GPT agents to within 3 percentage points.

\paragraph{Recent frontier models reach nearly 100\% compilation accuracy, shifting the bottleneck to test execution.} Across earlier model generations, a large fraction of generated patches fail to compile (Table~\ref{tab:main-results}), with compilation rates reaching at most 49.6\% among standard earlier models in the oracle setting (e.g., 48.3\% for GPT-5.2 and 10.5\% for Gemini 2.5 Pro) and 58.5\% in the agent setting. By contrast, the latest models—GPT-6-Astra, GPT-6.1-Sol, GPT-5.6-Sol, DeepSeek-V4.1-Flash, and DeepSeek-V4-Pro, along with GPT-5.5—produce applicable patches on 94.0\%--100.0\% of tasks and reach nearly 100\% compilation accuracy across both settings (100.0\% for GPT-6-Astra, GPT-6.1-Sol, and DeepSeek-V4.1-Flash). This sharp improvement across model generations reflects rapid progress in handling CUDA/C++ syntax, cross-file type consistency, and heterogeneous GPU build systems, largely eliminating build failures and leaving functional test verification as the primary bottleneck.

\subsection{Fine-Grained Breakdown}

\begin{figure}[!htb]
  \centering
  \includegraphics[width=\textwidth]{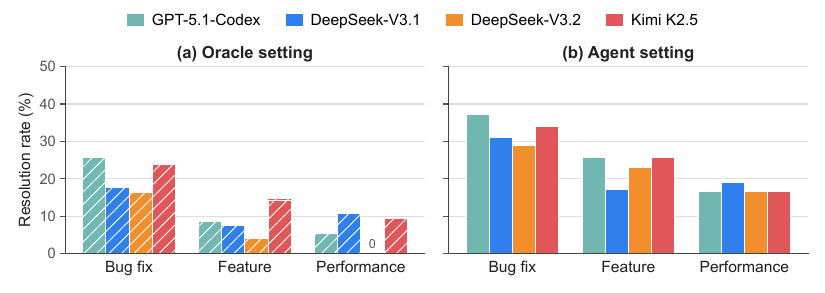}
  \caption{Resolution rate by task type in the oracle (a) and agent (b) settings, for four representative models evaluated in both settings. Appendix~\ref{app:breakdown} lists the values.}
  \label{fig:task-type-results}
\end{figure}

\paragraph{Feature and performance tasks have lower resolution rates than bug fixes.} In both settings, all four models shown in Figure~\ref{fig:task-type-results} resolve a smaller fraction of feature and performance tasks than bug fixes. Kimi K2.5, for instance, resolves 23.8\% of bug fixes versus 14.4\% of feature tasks in the oracle setting, and 34.0\% versus 25.6\% in the agent setting. Whereas bug fixes often reduce to localized boundary checks or numerical corrections, feature implementations and performance optimizations typically require adding new template instantiations, dispatch paths, or host bindings across multiple components.

\begin{figure}[!htb]
  \centering
  \includegraphics[width=\textwidth]{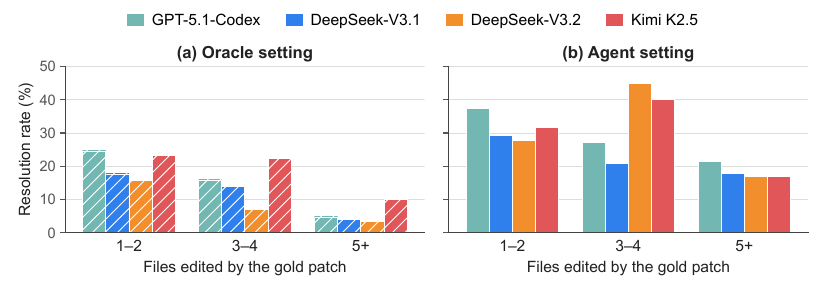}
  \caption{Resolution rate by the number of files that the gold patch edits, in the oracle (a) and agent (b) settings.}
  \label{fig:file-scope-results}
\end{figure}

\paragraph{Resolution rates drop on patches that modify five or more files.} In the oracle setting, resolution rates drop sharply on tasks that edit five or more files compared with tasks that edit one or two files, with DeepSeek-V3.1 dropping from 17.7\% to 4.0\% (Figure~\ref{fig:file-scope-results}). Without compiler feedback, single-pass edits across many files frequently introduce cross-file signature or type mismatches. In the agent setting, resolution rates also decline on the 5+ file bin (for example, GPT-5.1-Codex drops from 37.5\% on 1--2 files to 21.4\% on 5+ files, compared with 24.6\% to 4.8\% in the oracle setting).

\begin{figure}[!htb]
  \centering
  \includegraphics[width=\textwidth]{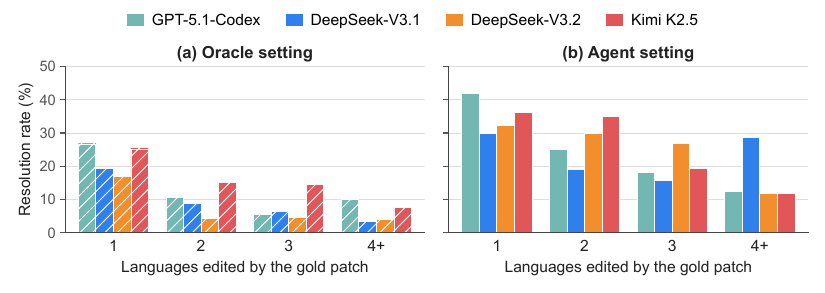}
  \caption{Resolution rate by the number of languages that the gold patch edits, in the oracle (a) and agent (b) settings.}
  \label{fig:language-results}
\end{figure}

\paragraph{Single-language tasks achieve higher resolution rates.} Across both settings, models generally achieve higher resolution rates on tasks that modify a single language than on multi-language tasks (Figure~\ref{fig:language-results}). In the oracle setting, resolution rates decline substantially from single-language tasks to tasks spanning four or more languages; in the agent setting, multi-language tasks—which require keeping device kernels, C++ host wrappers, Python interfaces, and build configurations synchronized—also score lower than single-language tasks.

\subsection{Perf-Replay Evaluation}

\begin{table}[!htb]
  \centering
  \small
  \caption{Agent results on the Perf-Replay subset. \textbf{Patch}, \textbf{Compile}, and \textbf{Test} report task counts, and \textbf{Perf.\ gain} reports the mean relative improvement of the target scalar metric over the base commit across the tasks that pass \textbf{Test}.}
  \label{tab:perf-replay}
  \begin{tabular}{@{}lrrrr@{}}
    \toprule
    Model & Patch & Compile & Test & Perf.\ gain (\%) \\
    \midrule
    Gold patch                 & 39/39 & 39/39 & 39/39 & 6.4 \\
    \midrule
    GPT-5.1-Codex (Codex)      & 39/39 & 15/39 & 3/39 & 3.3 \\
    GPT-5-Codex (Codex)        & 39/39 & 10/39 & 2/39 & 2.4 \\
    Kimi K2.5 (OpenCode)       & 35/39 &  8/39 & 2/39 & 3.3 \\
    DeepSeek-V3.2 (OpenCode)   & 39/39 & 12/39 & 1/39 & 4.3 \\
    \bottomrule
  \end{tabular}
\end{table}

\paragraph{Passing the correctness gate remains the primary bottleneck on Perf-Replay.} On the Perf-Replay subset, evaluated agents generate applicable patches for most tasks, but many fail during compilation and only a small fraction pass the functional correctness tests $T$ (Table~\ref{tab:perf-replay}). Because each agent resolves only a few tasks—and not necessarily the same subset of tasks—the resulting \textbf{Perf.\ gain} values are computed over small, model-specific passing sets and should be interpreted as preliminary observations rather than direct cross-model speedup rankings. On the tasks they do resolve, candidate patches achieve modest relative improvements over the base commit; Appendix~\ref{app:perf-cases} examines representative case studies across models to illustrate how workload shapes, metric definitions, and patch implementations affect measured speedups.

\section{Related Work}
\label{sec:related-work}

\begin{table}[!htb]
  \centering
  \small
  \caption{Comparison of SWE-GPU-Bench with representative GPU kernel and repository-level software engineering benchmarks. \textbf{Repo-level}: tasks require editing an existing multi-file repository; \textbf{Real PRs}: tasks are derived from merged pull requests or commits; \textbf{Multi-lang.\ \& build}: tasks span coordinated edits across device code, host bindings, and build configurations (e.g., CUDA/C++/Python/CMake); \textbf{GPU exec.}: patches are compiled and validated on GPU hardware.}
  \label{tab:related}
  \setlength{\tabcolsep}{3.5pt}
  \begin{tabularx}{\textwidth}{@{}lcccc>{\raggedright\arraybackslash}X@{}}
    \toprule
    Benchmark & Repo-level & Real PRs & Multi-lang.\ \& build & GPU exec. & Task scope \\
    \midrule
    KernelBench~\citep{kernelbench}             & --         & --         & --         & \checkmark & Isolated kernel generation \\
    TritonBench~\citep{tritonbench}             & --         & --         & --         & \checkmark & Isolated operator generation \\
    FlashInfer-Bench~\citep{flashinferbench}    & --         & --         & --         & \checkmark & Kernel \& inference workflow \\
    \midrule
    SWE-bench~\citep{swebench}                  & \checkmark & \checkmark & --         & --         & Issue resolution (Python) \\
    SWE-bench-Live~\citep{swebenchlive}         & \checkmark & \checkmark & --         & --         & Live issue resolution (Python) \\
    SWE-Perf~\citep{sweperf}                    & \checkmark & \checkmark & --         & --         & Performance optimization (Python) \\
    GSO~\citep{gso}                             & \checkmark & \checkmark & \checkmark & --         & Performance optimization (CPU) \\
    \midrule
    ParEval-Repo~\citep{parevalrepo}            & \checkmark & --         & \checkmark & \checkmark & Parallel model translation \\
    ISO-Bench~\citep{isobench}                  & \checkmark & \checkmark & --         & \checkmark & Inference engine optimization \\
    \textbf{SWE-GPU-Bench (ours)}               & \checkmark & \checkmark & \checkmark & \checkmark & \textbf{Bug fix, feature, performance} \\
    \bottomrule
  \end{tabularx}
\end{table}

\paragraph{GPU kernel benchmarks.} KernelBench, TritonBench, CUDABench, MultiKernelBench, and ComputeEval evaluate generated CUDA or Triton kernels for functional correctness and runtime performance~\citep{kernelbench,tritonbench,cudabench,multikernelbench,computeeval}. TritonGym, robust-kbench, GEAK, FlashInfer-Bench, and SOL-ExecBench incorporate agentic refinement, inference workloads, or hardware-limit targets~\citep{tritongym,robustkbench,geak,flashinferbench,solexecbench}. These benchmarks evaluate individual kernels or operators in isolation, abstracting away the host bindings, build systems, and cross-file dependencies required to integrate changes into a full repository.

\paragraph{Repository-level software engineering benchmarks.} RepoBench, CrossCodeEval, and ReCUBE evaluate code completion with repository context~\citep{repobench,crosscodeeval,recube}, while work on repository-level reasoning highlights the difficulty of cross-file retrieval~\citep{repograph}. SWE-bench, SWE-Perf, FEA-Bench, OmniCode, and Rust-SWE-bench evaluate issue resolution, performance optimization, and feature implementation in real repositories~\citep{swebench,sweperf,feabench,omnicode,rustswebench}, commonly using agent frameworks such as SWE-agent or localization pipelines such as Agentless~\citep{sweagent,agentless}. To overcome static dataset contamination and manual environment curation in SWE-bench, SWE-bench-Live~\citep{swebenchlive} introduces a continuously updated Python issue-resolution benchmark built with RepoLaunch, an agentic pipeline that automatically provisions per-instance containers and validates fail-to-pass tests. While SWE-GPU-Bench similarly uses a coding agent to synthesize and replay container-level setup and test commands from real PRs, it targets the GPU software stack rather than CPU-bound Python packages—requiring multi-language source builds (CUDA, C++, Python, and CMake/Bazel) and architecture-specific execution on GPU hardware—and extends evaluation beyond bug fixes to feature implementations and correctness-gated performance optimizations. Similarly, GSO includes repositories that contain GPU code, but its released tasks execute on CPUs~\citep{gso}, whereas SWE-GPU-Bench compiles and validates patches on GPUs.

\paragraph{Repository-level GPU benchmarks.} Table~\ref{tab:related} compares SWE-GPU-Bench with representative kernel-level and repository-level benchmarks. ParEval-Repo evaluates code translation of HPC applications across parallel programming models (CUDA, OpenMP, and Kokkos), identifying build systems and cross-file dependencies as primary bottlenecks~\citep{parevalrepo}. ISO-Bench evaluates performance optimization commits drawn from LLM inference engines (vLLM and SGLang)~\citep{isobench}. Complementing these efforts, SWE-GPU-Bench targets everyday PR-driven maintenance across diverse layers of the GPU software stack, combining multi-language repository edits and source-level GPU builds across bug fixes, feature implementations, and correctness-gated performance optimizations.

\section{Discussion}
\label{sec:discussion}

\paragraph{Limitations and future directions.} Several limitations shape the scope of our findings. First, because the setup commands $S$ are executed before the agent starts, SWE-GPU-Bench decouples code maintenance from environment provisioning; evaluating whether agents can autonomously configure GPU toolchains and resolve system dependencies from scratch is an important direction for future work. Second, to ensure isolatable fail-to-pass test oracles, our filtering pipeline bounds patch size and excludes broad architectural refactors and release merges, scoping the benchmark to PR-scale maintenance. Third, all tasks are validated on NVIDIA A100 and H100 GPUs and have not yet been validated on newer architectures such as Blackwell or non-NVIDIA accelerators; tasks targeting specific compute capabilities (such as \texttt{sm\_80} or \texttt{sm\_90a}) and all runtime measurements require re-validation when ported to other hardware. Fourth, standardized performance evaluation across heterogeneous GPU workloads remains an open challenge: our Perf-Replay subset relies on task-specific benchmark commands that output scalar metrics on a relatively small set of tasks, supporting preliminary observations rather than definitive speedup rankings. Finally, problem statements are distilled by an LLM prior to manual audit, and each model configuration is evaluated in a single pass@1 run; Appendix~\ref{app:limitations} discusses additional considerations including pretraining contamination, test adequacy, and broader impacts.

\paragraph{Conclusion.} We introduced SWE-GPU-Bench, a benchmark for evaluating language models and coding agents on repository-level GPU software engineering across CUDA, C++, Python, and heterogeneous build systems. By pairing real-world pull requests with container-verified setup, test, and performance benchmark commands, SWE-GPU-Bench measures end-to-end capabilities from cross-file localization and compilation to functional verification and correctness-gated speedup evaluation. While recent frontier models have advanced rapidly to reach nearly 100\% compilation accuracy—overcoming the build failures that hindered earlier generations—the strongest agent configurations resolve up to 58.3\% of tasks under our pass@1 evaluation protocol, leaving functional test verification and the correctness gate on performance tasks as major bottlenecks. We release the benchmark dataset, construction and evaluation pipeline, and container specifications to support future research on GPU coding agents.

\begin{ack}
The authors received no specific third-party funding for this work and declare no competing financial interests.
\end{ack}

\bibliographystyle{plainnat}
\bibliography{neurips_2026}

\clearpage
\appendix

\section{Benchmark Details}
\label{app:benchmark-details}

This appendix gives the details of the construction pipeline in Section~\ref{sec:construction}.

\subsection{Collection and Filtering}
\label{app:filtering}

\paragraph{Repositories and PRs.} We identify candidate repositories through GPU-related GitHub searches and CUDA-related repository signals, and retain those actively maintained by an official organization or by core maintainers. For each repository, we collect its merged PRs with their title, description, metadata, changed files, diff, and any available CI information.

\paragraph{Rule-based filter.} A PR passes this filter if it modifies at least one test file and at least one GPU-relevant implementation file (CUDA, C++, Triton, ROCm, GPU runtime or backend code, kernel dispatch, or build and compilation files). We exclude PRs that modify only documentation, tests, release metadata, or pure Python code, and bound the number of modified files and lines across the full textual diff (distribution in Table~\ref{tab:percentiles}). Bounding diff size excludes release merges and broad refactors that bundle multiple unrelated changes and cannot be reduced to a single task with an isolatable fail-to-pass test; SWE-GPU-Bench therefore targets PR-scale maintenance and leaves release engineering and architectural migrations out of scope.

\paragraph{LLM judge.} The judge is \texttt{gpt-5.2-2025-12-11}, which we query in a single chat call without an agent harness. It places each candidate PR in one primary category (bug fix, performance, feature, compatibility, observability, build infrastructure, refactor, test-only, documentation-only, release, or mixed) and rates its GPU relevance, problem clarity, test linkage, and reproducibility on A100 or H100 GPUs. A PR advances only if the judge accepts it or marks it as borderline, it changes core code, and it is rated reproducible. Its GPU relevance and problem clarity must also not be rated low, and its tests must be strongly linked to the change, or moderately linked with a clear oracle. Before freezing the benchmark, we refined the judge prompt on observed failure cases. Appendix~\ref{app:prompt-filter} describes the prompt.

\paragraph{Verification and final composition.} Candidates passing the rule-based and LLM filters proceed to interactive command synthesis (Stage~III), clean fail-to-pass replay in an agent-free container (Stage~IV), and manual audit (Stage~V), yielding $417$ verified tasks across $30$ repositories: $378$ general-set tasks and $39$ Perf-Replay tasks (spanning $399$ unique pull requests, with $18$ instances shared across both tracks).

\subsection{Task Format}
\label{app:format}

Each task is a single JSON record with the fields in Table~\ref{tab:fields}. The listing below shows the problem statement of CUTLASS PR~\#1275, whose synthesized commands appear in Appendix~\ref{app:case-study}.

\begin{lstlisting}[style=promptstyle]
Extend the GEMM epilogue "tensor broadcast" bias feature to support broadcasting a 1D bias vector either per output row (existing behavior) or per output column, selectable at compile time. Preserve the current default behavior so existing code and kernels continue to treat the bias as per-row unless explicitly opted into per-column mode.

Also ensure the host-side reference epilogue computation used by utilities/tests applies the bias consistently with the selected broadcast direction, so correctness checks match device behavior.

Constraints:
- Do not modify, add, or update any test sources or golden outputs.
- Keep public APIs/backward compatibility: existing template instantiations and callers must compile unchanged.
- The per-column option must work with the existing combination of scaling (alpha/beta), bias binary op, and activation/unary/binary epilogue features.

Success criteria:
- Repository builds successfully with CUDA targets enabled.
- Existing unit tests pass unchanged.
- A kernel configured to use per-column bias produces correct results (matching the host reference) across typical MxN shapes and batch/L dimensions.
\end{lstlisting}

\begin{table}[!htb]
  \centering
  \small
  \caption{Fields of a SWE-GPU-Bench task.}
  \label{tab:fields}
  \begin{tabularx}{\textwidth}{@{}lX@{}}
    \toprule
    Field & Description \\
    \midrule
    \texttt{instance\_id} & Unique identifier, \texttt{owner\_\_repo\_\_PR}. \\
    \texttt{repo}, \texttt{pr\_number} & Source repository and pull request number. \\
    \texttt{base\_commit}, \texttt{head\_commit} & Commit before the pull request and its merged head. \\
    \texttt{problem\_statement} & Task description shown to the model. \\
    \texttt{task\_type} & Bug fix, feature, or performance. \\
    \texttt{setup\_commands} & Commands that provision the environment. \\
    \texttt{test\_commands} & Commands that compile and run the tests of the pull request. \\
    \texttt{patch} & Gold patch $\Delta_{\mathrm{gold}}$ (hidden from the model). \\
    \texttt{test\_patch} & Test patch $\Delta_{\mathrm{test}}$ (hidden from the model). \\
    \texttt{perf\_commands} & Performance commands $P$ (Perf-Replay tasks only). \\
    \texttt{docker\_image} & Base container image. \\
    \bottomrule
  \end{tabularx}
\end{table}

\subsection{Size Distributions}
\label{app:stats}

Patch sizes are heavy-tailed (Table~\ref{tab:percentiles}): the median gold patch changes 80 lines in 3 files, while the largest gold patches reach 3{,}227 lines and 62 files.

\begin{table}[!htb]
  \centering
  \small
  \caption{Distribution of size attributes over all tasks.}
  \label{tab:percentiles}
  \setlength{\tabcolsep}{4pt}
  \begin{tabular}{@{}lrrrrrrrrr@{}}
    \toprule
    Attribute & Mean & SD & P25 & P50 & P75 & P90 & P95 & P99 & Max \\
    \midrule
    Gold-patch lines  & 178.6 & 308.6 & 21 & 80 & 200 & 415.2 & 631.8 & 1{,}339.9 & 3{,}227 \\
    Test-patch lines  & 126.0 & 148.9 & 33.5 & 76 & 155.5 & 328.6 & 433.6 & 705.2 & 927 \\
    Gold-patch files  & 3.7 & 4.4 & 1 & 3 & 5 & 7 & 9 & 19 & 62 \\
    Problem words     & 236.6 & 237.2 & 132 & 190 & 261 & 427.2 & 621.2 & 1{,}347.2 & 2{,}636 \\
    \bottomrule
  \end{tabular}
\end{table}

\subsection{Manual Review}
\label{app:human-validation}

Every candidate task that passes replay validation is manually audited.

\paragraph{Problem statement.} The problem statement must describe the intended change accurately without revealing implementation details from the gold patch. An automated script first flags tokens in the statement that also appear in the gold patch, including file paths, function names, internal identifiers, and code fragments. Reviewers inspect each flagged token—retaining public API names required by the tests or stated in the original issue—and revise or reject any statement that discloses edit locations, introduces unsupported requirements, omits core requirements, or requests test modifications.

\paragraph{Test commands.} Although replay validation ensures that $T$ fails before applying the gold patch and passes afterward, reviewers additionally verify that $T$ directly exercises the test files, functions, markers, or binaries introduced in $\Delta_{\mathrm{test}}$. Commands that miss the modified tests or execute an unnecessarily broad test suite are revised or rejected.

\paragraph{Performance commands.} For Perf-Replay tasks, reviewers verify that $P$ executes the benchmark relevant to the PR and that the evaluation targets the primary metric (latency, throughput, or kernel time) corresponding to the PR's optimization goal.

\subsection{Example: Synthesized Commands for CUTLASS PR \#1275}
\label{app:case-study}

CUTLASS PR~\#1275 (``Allow per-column bias in EpilogueTensorBroadcast'', +71/$-$9 lines across 4 files) adds a compile-time per-column bias option alongside an SM90 GoogleTest unit test. Because the PR description does not specify build or test instructions, the command-synthesis agent located the CUDA compiler, configured CMake for the \texttt{sm\_90a} target architecture, compiled the CUTLASS test harness and the modified test source, linked against the CUDA runtime and driver stubs, and filtered execution to the new GoogleTest case. In the listing below, highlighted fragments depend on the environment or target architecture, and \texttt{\$INC} abbreviates repeated include paths.

\begin{lstlisting}[style=cmdstyle]
# setup: configure the build for SM90a
export CUDACXX=@H/usr/local/cuda/bin/nvccH@
cmake -S repo -B artifacts/build @H-DCUTLASS_NVCC_ARCHS=90aH@ -DCUTLASS_ENABLE_TESTS=ON \
      -DCUTLASS_ENABLE_CUBLAS=OFF -DCUTLASS_ENABLE_CUDNN=OFF
# setup: compile GoogleTest and the CUTLASS test harness
c++ -O3 -std=c++17 $INC @H-isystem /usr/local/cuda/includeH@ -c .../gtest-all.cc
c++ -O3 -std=c++17 -DCUTLASS_TEST_LEVEL=0 $INC -c repo/test/unit/common/filter_architecture.cpp
c++ -O3 -std=c++17 -DCUTLASS_TEST_LEVEL=0 $INC -c repo/test/unit/test_unit.cpp
# test: compile the changed test for SM90a
@H/usr/local/cuda/bin/nvccH@ -O3 -std=c++17 @H--generate-code=arch=compute_90a,code=[sm_90a]H@ \
      @H-DCUTLASS_ENABLE_TENSOR_CORE_MMA=1H@ --expt-relaxed-constexpr $INC \
      -c repo/test/unit/gemm/device/sm90_gemm_f32_f32_f32_tensor_op_f32_tensor_broadcast.cu
# test: link against the CUDA runtime and driver stubs
c++ *.o -o artifacts/manual_build/percol_bias_test @H-L/usr/local/cuda/lib64/stubsH@ \
      @H/usr/local/cuda/lib64/libcudart.soH@ @H/usr/local/cuda/lib64/stubs/libcuda.soH@ -lpthread -ldl
# test: run only the new test case
./artifacts/manual_build/percol_bias_test \
  --gtest_filter=SM90_Device_Gemm_f32t_f32n_f32n_tensor_op_gmma_f32_tensor_broadcast.@H*_PerColBiasH@
\end{lstlisting}

\section{Experimental Details}
\label{app:experimental-details}

\subsection{Models and Harnesses}
\label{app:models}

Table~\ref{tab:models} lists the evaluated models and their versions. In the oracle setting, each model is queried once with the prompt in Appendix~\ref{app:prompt-oracle}. In the agent setting, GPT models run in Codex CLI and the other models in OpenCode, with the prompt in Appendix~\ref{app:prompt-agent}.

\paragraph{Dataset revision and evaluation denominators.} During the camera-ready revision, we audited and updated the benchmark dataset by removing invalid tasks and refining task definitions and verification scripts. Recent frontier models—specifically GPT-5.6-Sol and subsequent models (GPT-6-Astra, GPT-6.1-Sol, DeepSeek-V4.1-Flash, and DeepSeek-V4-Pro)—were evaluated across the full updated general benchmark set. Because the API endpoints for several earlier models were no longer accessible at the time of the revision, these earlier baselines could not be re-run on newly added or updated tasks; their metrics in Table~\ref{tab:main-results} (and the fine-grained breakdowns in Section~\ref{sec:results} and Appendix~\ref{app:breakdown}) are therefore computed over the retained subset of valid tasks from their prior evaluation runs, resulting in different evaluation denominators across model generations.

\begin{table}[!htb]
  \centering
  \small
  \caption{Evaluated models. A dash in the harness column marks a model run only in the oracle setting.}
  \label{tab:models}
  \begin{tabular}{@{}llll@{}}
    \toprule
    Model & Version or endpoint & Settings & Agent harness \\
    \midrule
    GPT-6.1-Sol         & \texttt{gpt-6.1-sol}                 & Oracle, agent  & Codex CLI \\
    GPT-6-Astra         & \texttt{gpt-6-astra}                 & Oracle, agent  & Codex CLI \\
    GPT-5.6-Sol         & \texttt{gpt-5.6-sol}                 & Oracle, agent  & Codex CLI \\
    GPT-5.5             & \texttt{gpt-5.5}                     & Agent          & Codex CLI \\
    GPT-5.2             & \texttt{gpt-5.2-2025-12-11}          & Oracle         & -- \\
    GPT-5-Codex         & \texttt{gpt-5-codex}                 & Oracle, agent  & Codex CLI \\
    GPT-5.1-Codex       & \texttt{gpt-5.1-codex-2025-11-13}    & Oracle, agent  & Codex CLI \\
    GPT-5 mini          & \texttt{gpt-5-mini-2025-08-07}       & Agent          & Codex CLI \\
    Claude Sonnet 4     & \texttt{claude-sonnet-4}             & Oracle         & -- \\
    Gemini 2.5 Pro      & \texttt{gemini-2.5-pro-preview-05-06}& Oracle         & -- \\
    GLM-5               & \texttt{glm-5}                       & Oracle         & -- \\
    Kimi K2.5           & \texttt{kimi-k2.5}                   & Oracle, agent  & OpenCode \\
    DeepSeek-V4.1-Flash & \texttt{ds-v4.1-flash}               & Oracle, agent  & OpenCode \\
    DeepSeek-V4-Pro     & \texttt{ds-v4-pro}                   & Oracle, agent  & OpenCode \\
    DeepSeek-V4-Flash   & \texttt{ds-v4-flash}                 & Oracle, agent  & OpenCode \\
    DeepSeek-V3.1       & \texttt{deepseek-v3.1}               & Oracle, agent  & OpenCode \\
    DeepSeek-V3.2       & \texttt{deepseek-reasoner}           & Oracle, agent  & OpenCode \\
    DeepSeek-R1         & \texttt{deepseek-r1}                 & Oracle         & -- \\
    \bottomrule
  \end{tabular}
\end{table}

\subsection{Cost}
\label{app:cost}

\paragraph{Command synthesis.} Table~\ref{tab:synthesis-cost} summarizes the token usage and runtime of command synthesis with GPT-5.1-Codex for both setup-and-test commands and performance commands across candidate repositories.

\begin{table}[!htb]
  \centering
  \small
  \caption{Tokens and runtime per command-synthesis run.}
  \label{tab:synthesis-cost}
  \setlength{\tabcolsep}{4pt}
  \begin{tabular}{@{}lrrrrrrr@{}}
    \toprule
    Run type & Runs & Mean & Median & P90 & P95 & Max & Mean time \\
    \midrule
    Setup and test commands      & 1{,}259 & 4.83M & 3.46M & 10.14M & 13.83M & 32.35M & 17.0 min \\
    + performance commands       & 292     & 4.00M & 2.55M & 8.53M  & 13.53M & 29.08M & 14.4 min \\
    \bottomrule
  \end{tabular}
\end{table}

\paragraph{Model evaluation.} Table~\ref{tab:frontier-cost} reports the per-task token consumption and wall-clock runtime across both the agent and oracle settings for the five most recent frontier models on A100 and H100 GPUs (wall-clock runtime depends on API latency and serving load, so it reflects empirical measurements in our evaluation environment):
\begin{itemize}
  \item \textbf{Multi-turn agent interaction consumes $\sim\!64\times$ more tokens per task than single-pass oracle generation.} In the agent setting, tasks average $3{,}018{,}335$ total tokens ($2{,}988{,}657$ input and $29{,}678$ output tokens), compared with $46{,}937$ total tokens ($42{,}182$ input and $4{,}755$ output tokens) in the oracle setting, reflecting repeated context accumulation from repository traversal, build logs, and test feedback.
  \item \textbf{Container compilation and test grading account for 4--5 minutes per task in both settings.} In the oracle setting, single-pass model inference averages $1.14$~min out of $5.72$~min total task time, with the remaining $\sim\!4.6$~min spent on container setup, compilation, and test execution. In the agent setting, total task duration averages $20.81$~min, of which $16.51$~min is spent in the interactive agent execution stage and $\sim\!4.3$~min in final patch verification.
  \item \textbf{Frontier models differ substantially in token and time efficiency.} Among agent configurations, GPT-6-Astra and GPT-6.1-Sol achieve the highest resolution rates ($58.3\%$ and $58.0\%$) with the lowest token footprints ($2.43$M and $1.71$M total tokens per task) and shortest model execution times ($9.55$ and $6.50$~min). By contrast, DeepSeek-V4-Pro generates $61{,}174$ output tokens per task on average—more than four times the output volume of the GPT agents—resulting in a mean model execution time of $35.18$~min ($39.36$~min total) while resolving $43.3\%$ of tasks.
\end{itemize}

\begin{table}[!htb]
  \centering
  \footnotesize
  \caption{Per-task mean runtime and token usage for recent frontier models in the agent and oracle settings. \textbf{Total time} is the end-to-end task duration (including container grading), and \textbf{Model time} is the model/agent execution stage.}
  \label{tab:frontier-cost}
  \setlength{\tabcolsep}{4pt}
  \begin{tabular}{@{}llrrrrr@{}}
    \toprule
    Setting & Model & Total time (min) & Model time (min) & Input tokens & Output tokens & Total tokens \\
    \midrule
    \multirow{6}{*}{Agent}
      & GPT-6.1-Sol         & 11.26 &  6.50 & 1{,}699{,}679 & 12{,}932 & 1{,}712{,}611 \\
      & GPT-6-Astra         & 14.26 &  9.55 & 2{,}413{,}145 & 13{,}848 & 2{,}426{,}993 \\
      & GPT-5.6-Sol         & 15.66 & 11.00 & 3{,}480{,}624 & 14{,}327 & 3{,}494{,}950 \\
      & DeepSeek-V4.1-Flash & 16.31 & 12.80 & 3{,}589{,}341 & 32{,}996 & 3{,}622{,}337 \\
      & DeepSeek-V4-Pro     & 39.36 & 35.18 & 2{,}968{,}245 & 61{,}174 & 3{,}029{,}418 \\
      \cmidrule(l){2-7}
      & \emph{Mean}         & 20.81 & 16.51 & 2{,}988{,}657 & 29{,}678 & 3{,}018{,}335 \\
    \midrule
    \multirow{6}{*}{Oracle}
      & GPT-6.1-Sol         &  5.83 &  1.28 &    43{,}811 &  4{,}634 &    48{,}446 \\
      & GPT-6-Astra         &  6.09 &  1.86 &    43{,}139 &  5{,}607 &    48{,}746 \\
      & GPT-5.6-Sol         &  6.26 &  1.15 &    42{,}674 &  5{,}603 &    48{,}277 \\
      & DeepSeek-V4.1-Flash &  5.12 &  0.31 &    39{,}167 &  7{,}222 &    46{,}389 \\
      & DeepSeek-V4-Pro     &  4.78 &  0.46 &    39{,}898 &  1{,}786 &    41{,}684 \\
      \cmidrule(l){2-7}
      & \emph{Mean}         &  5.72 &  1.14 &    42{,}182 &  4{,}755 &    46{,}937 \\
    \bottomrule
  \end{tabular}
\end{table}

\section{Additional Results}
\label{app:breakdown}

Table~\ref{tab:breakdown} lists the values plotted in Figures~\ref{fig:task-type-results}--\ref{fig:language-results} for the four representative baseline models (computed over their prior evaluation runs as noted in Appendix~\ref{app:models}).

\begin{table}[!htb]
  \centering
  \scriptsize
  \caption{Resolution rate (\%) by task type and by the number of files and languages edited by the gold patch. Perf.: performance.}
  \label{tab:breakdown}
\begin{tabular}{@{}llrrrrrrrrrr@{}}
\toprule
 & & \multicolumn{3}{c}{Task type} & \multicolumn{3}{c}{Files edited} & \multicolumn{4}{c}{Languages edited} \\
\cmidrule(lr){3-5}\cmidrule(lr){6-8}\cmidrule(l){9-12}
Setting & Model & Bug fix & Feature & Perf. & 1--2 & 3--4 & 5+ & 1 & 2 & 3 & 4+ \\
\midrule
Oracle & GPT-5.1-Codex & 25.8 & 8.6 & 5.4 & 24.6 & 15.9 & 4.8 & 26.7 & 10.6 & 5.4 & 10.0 \\
Oracle & DeepSeek-V3.1 & 17.8 & 7.7 & 10.8 & 17.7 & 14.0 & 4.0 & 19.4 & 8.9 & 6.5 & 3.3 \\
Oracle & DeepSeek-V3.2 & 16.5 & 4.0 & 0.0 & 15.7 & 7.0 & 3.4 & 17.0 & 4.2 & 4.5 & 4.0 \\
Oracle & Kimi K2.5 & 23.8 & 14.4 & 9.4 & 23.2 & 22.2 & 9.9 & 25.2 & 15.2 & 14.5 & 7.7 \\
\midrule
Agent & GPT-5.1-Codex & 37.3 & 25.6 & 16.7 & 37.5 & 27.3 & 21.4 & 42.0 & 25.0 & 18.2 & 12.5 \\
Agent & DeepSeek-V3.1 & 31.0 & 17.2 & 18.9 & 29.3 & 20.8 & 17.9 & 30.0 & 19.2 & 15.7 & 28.6 \\
Agent & DeepSeek-V3.2 & 28.9 & 23.1 & 16.7 & 27.8 & 45.0 & 17.1 & 32.4 & 30.0 & 26.9 & 11.8 \\
Agent & Kimi K2.5 & 34.0 & 25.6 & 16.7 & 31.6 & 40.0 & 17.1 & 36.1 & 35.0 & 19.2 & 11.8 \\
\bottomrule
\end{tabular}

\end{table}

\subsection{Perf-Replay Case Studies}
\label{app:perf-cases}

Because performance tasks span heterogeneous metrics, workload shapes, and hardware bottlenecks, aggregate speedup numbers cannot be interpreted as a direct cross-model ranking. Even when a candidate patch passes all functional tests in $T$, its runtime behavior depends on low-level implementation choices relative to the reference gold patch. Table~\ref{tab:perf-cases} reports the normalized task-level performance gains for two representative agents in Table~\ref{tab:perf-replay} (GPT-5.1-Codex and Kimi K2.5), and Figures~\ref{fig:case-flashinfer}--\ref{fig:case-cutlass} contrast the generated code against the gold patch on three representative tasks.

\begin{table}[!htb]
  \centering
  \small
  \caption{Normalized task-level performance gains on representative Perf-Replay tasks passing functional tests, aligned with \textbf{Perf.\ gain (\%)} in Table~\ref{tab:perf-replay}.}
  \label{tab:perf-cases}
  \begin{tabular}{@{}lllrr@{}}
    \toprule
    Model & Task & Metric & Agent gain (\%) & Gold gain (\%) \\
    \midrule
    \multirow{3}{*}{GPT-5.1-Codex}
      & \texttt{cutlass\#53}        & Throughput &  0.3 &  0.1 \\
      & \texttt{flashinfer\#588}    & Latency    &  2.1 & 11.4 \\
      & \texttt{tensorflow\#113198} & Throughput &  7.5 &  7.7 \\
    \cmidrule(l){2-5}
    \multirow{2}{*}{Kimi K2.5}
      & \texttt{cuda-python\#459}   & Latency    &  6.5 &  6.3 \\
      & \texttt{cutlass\#53}        & Throughput &  0.1 &  0.1 \\
    \bottomrule
  \end{tabular}
\end{table}

\paragraph{Case 1: Kernel launch configuration and host--device synchronization (\texttt{flashinfer\#588}).}
This task replaces a batch-parallel paged KV-cache append kernel (\texttt{dim3 nblks(batch\_size)}, which under-utilizes the GPU at small batch sizes) with an element-parallel implementation. Both the gold patch and GPT-5.1-Codex convert ragged \texttt{append\_indptr} offsets into per-token \texttt{(batch\_indices, positions)} metadata and pass all functional tests, but their implementations differ in three key places (Figure~\ref{fig:case-flashinfer}): the gold patch precomputes metadata outside the timed loop, bounds grid size via occupancy query, and uses fast reciprocal division (\texttt{uint\_fastdiv}), whereas GPT-5.1-Codex recomputes metadata inside the timed region, triggers an implicit device-to-host synchronization via \texttt{.item<int32\_t>()}, and launches one block per token with standard integer division.

\begin{figure}[!htb]
  \centering
  \begin{minipage}[t]{0.488\textwidth}
    \noindent\colorbox{gold accent!15}{\parbox{\dimexpr\linewidth-2\fboxsep\relax}{\scriptsize\bfseries\color{gold accent} Gold Patch ($\Delta_{\mathrm{gold}}$: $+11.4\%$ gain)}}
    \begin{lstlisting}[style=goldcode]
// [1] Precompute metadata outside timed loop
batch_idx, pos = @Gget_batch_indices_positionsG@(...);

// [2] Occupancy-bounded grid-stride launch
@GcudaOccupancyMaxActiveBlocksPerMultiprocessorG@(
    &b_per_sm, kernel, num_threads, 0);
int n_cta = min(b_per_sm, ceil_div(nnz, sms)) * sms;
dim3 nblks(@Gn_ctaG@);

// [3] Fast reciprocal divmod in kernel loop
for (uint32_t i = cta_id; i < nnz; i += num_ctas) {
  paged_kv.page_size.@GdivmodG@(
      indptr[batch_idx[i]] * page_sz + pos[i],
      page_iter, entry_idx);
}
    \end{lstlisting}
  \end{minipage}\hfill
  \begin{minipage}[t]{0.488\textwidth}
    \noindent\colorbox{agent accent!15}{\parbox{\dimexpr\linewidth-2\fboxsep\relax}{\scriptsize\bfseries\color{agent accent} GPT-5.1-Codex Patch ($\hat{\Delta}$: $+2.1\%$ gain)}}
    \begin{lstlisting}[style=agentcode]
// [1] Recompute in timed loop + implicit D2H sync
int32_t total = @Aappend_indptr[-1].item<int32_t>()A@;

// [2] Unbounded launch: 1 CTA per token
//     (no occupancy query or grid-stride loop)
uint32_t num_tokens = append_key.size(0);
dim3 nblks(@Anum_tokensA@);

// [3] Standard integer division/modulo in kernel
uint32_t i = blockIdx.x;
if (i < num_tokens) {
  page_iter = indptr[batch_idx[i]]
            + @Aseq_pos[i] / page_sizeA@;
  entry_idx = @Aseq_pos[i] % page_sizeA@;
}
    \end{lstlisting}
  \end{minipage}
  \caption{Implementation comparison on \texttt{flashinfer-ai/flashinfer\#588}: (1) precomputed metadata vs.\ per-call device-to-host synchronization (\texttt{.item()}), (2) occupancy-bounded grid-stride loop vs.\ one CTA per token, and (3) fast reciprocal \texttt{divmod} vs.\ integer division.}
  \label{fig:case-flashinfer}
\end{figure}

\paragraph{Case 2: Bit-packed tagged pointers vs.\ double-word atomics (\texttt{tensorflow\#113198}).}
This task eliminates spin-waiting in XLA's \texttt{ObjectPool} by replacing a mark-bit spin lock with lock-free \texttt{PopEntry} and \texttt{PushEntry} operations while preventing ABA pointer reuse (Figure~\ref{fig:case-tensorflow}). The gold patch aligns each pool entry to a 64-byte cache line to avoid false sharing and packs a 13-bit ABA version tag into unused bits of a single 64-bit \texttt{uintptr\_t}, whereas GPT-5.1-Codex leaves \texttt{Entry} unaligned and pairs the pointer with a full 64-bit counter in a 16-byte \texttt{Head} struct requiring 128-bit double-word CAS.

\begin{figure}[!htb]
  \centering
  \begin{minipage}[t]{0.488\textwidth}
    \noindent\colorbox{gold accent!15}{\parbox{\dimexpr\linewidth-2\fboxsep\relax}{\scriptsize\bfseries\color{gold accent} Gold Patch ($\Delta_{\mathrm{gold}}$: $+7.7\%$ gain)}}
    \begin{lstlisting}[style=goldcode]
// [1] 64B cache-line aligned entry (no false sharing)
struct @Galignas(64)G@ Entry {
  std::optional<T> object;
  Entry* next = nullptr;
};
// [2] Pack 13-bit ABA tag into 64-bit pointer word
//     Layout: [7 hi tag bits | ptr | 6 lo tag bits]
std::atomic<@Guintptr_tG@> head_{0};

// [3] Single-word 64-bit atomic CAS
while (Entry* ptr = GetPtr(head)) {
  uintptr_t next = @GMakeTagged(ptr->next, head)G@;
  if (head_.compare_exchange_weak(head, next, ...))
    return std::unique_ptr<Entry>(ptr);
}
    \end{lstlisting}
  \end{minipage}\hfill
  \begin{minipage}[t]{0.488\textwidth}
    \noindent\colorbox{agent accent!15}{\parbox{\dimexpr\linewidth-2\fboxsep\relax}{\scriptsize\bfseries\color{agent accent} GPT-5.1-Codex Patch ($\hat{\Delta}$: $+7.5\%$ gain)}}
    \begin{lstlisting}[style=agentcode]
// [1] Unaligned entry (entries share cache lines)
struct @AEntryA@ {
  std::optional<T> object;
  Entry* next = nullptr;
};
// [2] Separate 64-bit counter in 16-byte struct
struct @Aalignas(16) Head { Entry* ptr; uintptr_t tag; }A@;
std::atomic<@AHeadA@> head_{Head{nullptr, 0}};

// [3] Double-word 128-bit atomic CAS
while (head.ptr != nullptr) {
  Head next{head.ptr->next, @Ahead.tag + 1A@};
  if (head_.compare_exchange_weak(head, next, ...))
    return std::unique_ptr<Entry>(head.ptr);
}
    \end{lstlisting}
  \end{minipage}
  \caption{Implementation comparison on \texttt{tensorflow/tensorflow\#113198}: (1) 64-byte cache-line alignment vs.\ default alignment, (2) 13-bit version tag packed into a 64-bit pointer vs.\ a 16-byte \texttt{Head} struct, and (3) 64-bit single-word CAS vs.\ 128-bit double-word CAS.}
  \label{fig:case-tensorflow}
\end{figure}

\paragraph{Case 3: Constructor-time vs.\ loop-level pointer disambiguation (\texttt{cutlass\#53}).}
In CUTLASS's Volta Tensor Core GEMM store iterator, indexing \texttt{pointer[ptr\_idx]} with a thread-dependent offset inside the unrolled store loop prevents NVCC from keeping the shared-memory pointers in registers (Figure~\ref{fig:case-cutlass}). Whereas the gold patch folds the warp-parity check into the constructor by swapping \texttt{pointer[0]} and \texttt{pointer[1]} once so that \texttt{store()} uses a compile-time constant index, both agents leave the constructor unchanged and instead eliminate dynamic array indexing inside \texttt{store()} via a precomputed register array (GPT-5.1-Codex) or a ternary select (Kimi K2.5).

\begin{figure}[!htb]
  \centering
  \begin{minipage}[t]{0.488\textwidth}
    \noindent\colorbox{gold accent!15}{\parbox{\dimexpr\linewidth-2\fboxsep\relax}{\scriptsize\bfseries\color{gold accent} Gold Patch ($\Delta_{\mathrm{gold}}$: $+0.1\%$ gain)}}
    \begin{lstlisting}[style=goldcode]
// [1] Swap pointers once in iterator constructor
if (((threadIdx.x >> 5) * Iterations::kD) & 2) {
  @GScalar *tmp = pointer[0];G@
  @Gpointer[0] = pointer[1]; pointer[1] = tmp;G@
}

// [2] Compile-time constant index in unrolled store()
CUTLASS_PRAGMA_UNROLL
for (int d = 0; d < Iterations::kD; ++d) {
  Scalar *_pointer = @Gpointer[(d & 1) ^ ((d >> 1) & 1)]G@;
  Store<...>::store(..., _pointer, offset);
}
    \end{lstlisting}
  \end{minipage}\hfill
  \begin{minipage}[t]{0.488\textwidth}
    \noindent\colorbox{agent accent!15}{\parbox{\dimexpr\linewidth-2\fboxsep\relax}{\scriptsize\bfseries\color{agent accent} Agent Patches ($\hat{\Delta}$: $+0.3\%$ / $+0.1\%$ gain)}}
    \begin{lstlisting}[style=agentcode]
// [1] GPT-5.1-Codex: precompute register array in store()
@AScalar *lane_ptrs[Iterations::kD];A@
for (int d = 0; d < Iterations::kD; ++d)
  lane_ptrs[d] = ptr_idx(d) ? pointer[1] : pointer[0];
Scalar *_pointer = @Alane_ptrs[d]A@;

// [2] Kimi K2.5: hoist warp_id + ternary select in store()
int warp_id = (threadIdx.x >> 5);
int ldg_idx = d + warp_id * Iterations::kD;
int ptr_sel = ((ldg_idx & 1) ^ ((ldg_idx >> 1) & 1));
Scalar *_pointer = @A(ptr_sel ? pointer[1] : pointer[0])A@;
// (both leave constructor unchanged)
    \end{lstlisting}
  \end{minipage}
  \caption{Implementation comparison on \texttt{NVIDIA/cutlass\#53}: the gold patch resolves warp-dependent pointer order once in the constructor so the unrolled \texttt{store()} loop uses compile-time constant indices, whereas the agents eliminate dynamic indexing locally inside \texttt{store()}.}
  \label{fig:case-cutlass}
\end{figure}

\section{Limitations and Broader Impact}
\label{app:limitations}

\paragraph{Limitations.} In addition to the limitations discussed in Section~\ref{sec:discussion}, SWE-GPU-Bench is drawn from a finite set of open-source repositories and does not span every GPU software stack or accelerator family. We exclude tasks that require non-NVIDIA accelerators at runtime, private infrastructure, or external services. Because tasks are constructed from public GitHub PRs, some repositories and commits may overlap with model pretraining corpora, and we have not quantified pretraining contamination. Moreover, passing the PR-derived tests verifies the behavior covered by $\Delta_{\mathrm{test}}$, which may not exhaustively test all edge cases or downstream invariants of the repository.

\paragraph{Broader impact.} SWE-GPU-Bench provides an empirical benchmark for evaluating and developing coding agents on GPU software, which underpins modern machine-learning training and inference infrastructure. Capable coding agents could reduce maintenance overhead across complex GPU codebases; however, automated edits to performance-critical kernels and runtime code risk introducing subtle numerical discrepancies or performance regressions, so human review and comprehensive testing remain essential before deployment. The benchmark is constructed exclusively from public open-source repositories, contains no personal information beyond public repository metadata, and executes candidate patches inside isolated containers.

\section{Prompts}
\label{app:prompts}

This appendix documents the prompts used to construct and evaluate SWE-GPU-Bench. The Stage~V problem-statement generation prompt (Appendix~\ref{app:prompt-statement}) and the evaluation prompts for the oracle and agent settings (Appendices~\ref{app:prompt-oracle} and~\ref{app:prompt-agent}) are reproduced verbatim. Because the prompts for Stage~II PR filtering (\texttt{prompts/fuse\_filter.md}) and Stage~III command synthesis (\texttt{prompts/agent\_explore.md}, and \texttt{prompts/agent\_perf\_repro.md} for performance commands) are multi-page operational specifications, we summarize their criteria and output schemas below and include the full templates in the \texttt{prompts/} directory of the code release.

\subsection{PR Filtering Prompt}
\label{app:prompt-filter}

The PR-filtering prompt specifies a rubric for the LLM judge to evaluate whether a merged PR is suitable as a GPU benchmark instance. Given the PR title, description, metadata, diff, modified file list, test diff, and available CI logs, the judge evaluates the candidate based solely on evidence present in these artifacts.

\textbf{Required judgments.} Each PR is assigned a primary category, optional secondary categories, and a scope level (\texttt{single\_file}, \texttt{single\_module}, \texttt{multi\_module}, or \texttt{repo\_level}). The rubric evaluates whether the PR modifies core GPU-relevant code, whether the accompanying tests exercise the modified code path, the strongest visible validation level, hardware reproducibility on A100 or H100 GPUs, and overall ratings for GPU relevance, problem clarity, test signal, and benchmark fit.

\textbf{Categories.} Allowed primary categories are \texttt{bugfix}, \texttt{perf}, \texttt{feature}, \texttt{compatibility}, \texttt{observability}, \texttt{build\_infra}, \texttt{refactor}, \texttt{test\_only}, \texttt{docs\_only}, \texttt{chore\_release}, and \texttt{mixed}. When a PR spans multiple aspects, the rubric prioritizes behavioral changes over maintenance categories according to the order \texttt{bugfix} $>$ \texttt{perf} $>$ \texttt{feature} $>$ \texttt{compatibility} $>$ \texttt{observability} $>$ \texttt{build\_infra} $>$ \texttt{refactor} $>$ \texttt{test\_only} $>$ \texttt{docs\_only} $>$ \texttt{chore\_release}.

\textbf{Hardware reproducibility.} The judge assesses whether the PR is reproducible on A100, H100, both, or neither (or marks \texttt{unclear} when evidence is insufficient). Candidates reproducible on A100 or H100 GPUs are retained, whereas PRs requiring unavailable hardware, non-NVIDIA runtime execution, or external services are rejected.

\textbf{Benchmark-fit rule.} Acceptance requires non-low GPU relevance and problem clarity, a substantive core-code modification, coupled test changes with strong test linkage (or medium linkage paired with a clear test oracle), hardware reproducibility on A100/H100, and a plausible local replay path. Candidates that satisfy most criteria but have a minor weakness on one dimension are labeled \texttt{borderline}; candidates lacking GPU relevance, core-code changes, test validation, or compatible hardware are labeled \texttt{reject}.

\textbf{Output format.} The judge outputs a JSON object conforming to the following schema:

\begin{lstlisting}[style=promptstyle]
{
  "pr_category": {
    "primary": "",
    "secondary": []
  },
  "core_code_change": false,
  "has_test_change": false,
  "test_link_strength": "",
  "code_test_coupled": false,
  "scope_level": "",
  "validation_level": "",
  "gpu_relevance": "",
  "problem_clarity": "",
  "test_signal": "",
  "hardware_reproducibility": "",
  "benchmark_fit": "",
  "reason": [
    "",
    "",
    ""
  ],
  "main_risk": "",
  "task_brief": ""
}
\end{lstlisting}

Each entry in \texttt{reason} must cite concrete evidence from the PR artifacts; when evidence is insufficient, the rubric requires outputting \texttt{borderline} or \texttt{reject}.

\subsection{Problem Statement Generation Prompt}
\label{app:prompt-statement}

\begin{lstlisting}[style=promptstyle]
<query_generator>
You are a task distillation assistant.

Your job is to convert GitHub pull request metadata and code patch information into a single English task prompt for a coding model.

The coding model will be given the repository itself, but it must NOT be given implementation spoilers from the patch.

Your goal is to produce a prompt that describes:
1. what needs to be achieved,
2. what constraints must be respected,
3. what counts as success,

while hiding patch-specific solution details.

## Input
You will receive:
- PR title
- PR body
- optional issue/context text
- the required interface: names the PR introduces that the hidden tests call
- patch/diff

## Objective
Write one concise but clear English prompt that instructs a coding model to reproduce the intent of the PR in the repository.

## Hard constraints
Do NOT reveal or mention:
- file paths
- file names
- function names
- class names
- identifiers
- line numbers
- exact code snippets
- exact strings from the patch unless absolutely necessary for understanding the task
- specific edit strategy
- the sequence of modifications
- how many places were changed

Do NOT summarize the patch mechanically.
Do NOT say "modify X in file Y".
Do NOT expose the answer.

## Required interface exception
The hidden tests call every name listed under "Required interface" below. A coding model cannot guess
these names, so the prompt MUST mention each of them, as the public function, class, parameter or
binding that has to exist, and say briefly what it is for. This is the only exception to the
identifier rule above. Do not reveal how they are implemented. If the list says "(none)", mention
no identifiers.

## What the output prompt SHOULD contain
The prompt should:
- describe the user-visible or maintainer-visible goal of the change
- state any important scope restrictions
- preserve behavior unless the PR is intended to change behavior
- include acceptance criteria that can be checked from the repository state
- be phrased as an implementation task for a coding model

## Style requirements
- Output English only
- Output only the final prompt text, no explanation
- Keep it concrete, actionable, and repository-aware
- Prefer task intent over patch mechanics
- If the PR is small or localized, keep the prompt concise
- If the PR changes behavior, describe expected behavior and edge cases, but not the exact implementation

## Compression rule
Abstract from implementation details to task intent:
- patch detail -> behavioral/task requirement
- file-level evidence -> subsystem-level scope
- exact edits -> acceptance criteria

## Domain wording
Describe the change in the terms the PR itself supports. Many PRs in these repositories touch GPU
code (kernels, CUDA or HIP code paths, GPU-facing APIs, tensor/layout semantics, memory access or
synchronization, numerical correctness), but others are compiler passes, build logic, host-side
runtime or Python API changes. Use GPU wording only when the PR is actually about GPU behavior; do
not reframe other changes as GPU work, and do not invent performance, kernel, or numerical
requirements that the PR content does not support.

## Examples of good abstraction in GPU repositories
Bad: "Change the launch configuration in kernel_x to use 256 threads"
Good: "Correct the GPU execution configuration issue so that the affected code path behaves as intended under the expected workload, while preserving surrounding interfaces, numerical behavior, and compatibility constraints."

Bad: "Add a guard in the CUDA helper when stride == 0"
Good: "Fix the edge-case handling in the affected GPU utility logic so that invalid or degenerate inputs are handled correctly without introducing regressions in normal execution paths."

## Output checklist
Before producing the final prompt, ensure:
- it does not leak implementation details from the patch
- it names every entry of the required interface, and no other identifiers
- it is specific enough for a coding model to act on
- it reflects the actual intent of the PR
- it includes sensible constraints against unrelated edits

Very IMPORTANT:
Assume test patches are handled separately: the coding model must implement the task without modifying, adding, or updating any test files, fixtures, snapshots, or expected outputs.

Now read the PR information and generate the final prompt.

PR repo:
{repo}

PR title:
{title}

PR body:
{body}

Additional context:
{context}

Required interface:
{interface}

Patch:
{patch}
</query_generator>
\end{lstlisting}

\subsection{Command Synthesis Prompt}
\label{app:prompt-commands}

The Stage~III command-synthesis prompt (\texttt{prompts/agent\_explore.md}) instructs the coding agent to construct reproducible setup and test commands (and \texttt{prompts/agent\_perf\_repro.md} analogously synthesizes performance benchmark commands). The prompt requires the agent to operate strictly as a reproducibility verifier—prohibiting source-code modifications, unverified test claims, or skipped steps—and to return a structured JSON verdict containing \texttt{setup\_commands}, \texttt{test\_commands}, and execution logs.

\textbf{Objective and input.} Given a JSON record containing repository metadata, PR number, title, body, base and head commit SHAs, diff text, and commit messages, the agent inspects the checked-out repository at \texttt{head\_sha}, provisions the environment, identifies and executes the smallest test target that covers the PR's changes, and records the resulting command lists.

\textbf{Environment and build policy.} The agent inspects the repository's build scripts, CI configurations, and documentation to configure Python packages and native dependencies—including CUDA, NVCC, CMake, Ninja, GCC/Clang, NCCL, and LLVM—matching the target repository and GPU architecture.

\textbf{Test-selection priority.} Candidate test targets are prioritized as follows: (1) tests added or modified by the PR, (2) tests directly exercising the modified modules, (3) the closest CI test target covering the modified code path, and (4) a minimal file- or module-level test target. Running an unrelated full test suite is prohibited.

\textbf{Hardware and system requirements.} For PRs modifying CUDA or Triton kernels, C++/CUDA extensions, custom operators, compiler passes, or distributed runtimes, the agent verifies whether execution requires specific GPU hardware, CUDA toolkit versions, compute capabilities, or native libraries, and marks the candidate as blocked when required hardware or system dependencies are unavailable in the container.

\subsection{Oracle Setting Prompt}
\label{app:prompt-oracle}

\begin{lstlisting}[style=promptstyle]
<oracle>
You are evaluating a benchmark instance with a fixed oracle context.

Use only the provided task description and code context.
Do not assume access to any files beyond the supplied context.
Return a minimal patch that fixes the task.

You will be provided with a partial code base and a task statement
explaining a problem to resolve.
<task>
{{problem_statement}}
</task>

The code context below is split into file blocks:
- `[start of path]` and `[end of path]` mark the exact file path for each block.
- Lines are prefixed as `0001| source text` for reference only.
- The `0001|` prefixes are not part of the files. Do not include them in the patch.
- Use the reference line numbers only to locate the target code regions.

<code>
{{golden-patch modified files}}
</code>

Return exactly one JSON object and nothing else. Do not return a unified diff.
The JSON object must have this shape:

```json
{
  "edits": [
    {
      "path": "relative/path/to/file.py",
      "old": "exact existing text block to replace",
      "new": "replacement text block"
    }
  ]
}
```

Rules:
- `path` must be one of the provided file paths unless you are adding a new helper file.
- `old` must be copied exactly from the file content, without the `0001|` line prefixes.
- For an existing file, `old` must match exactly once in that file.
- To add a new file, use `"old": ""` and put the complete file content in `new`.
- Do not include code-context line prefixes such as `0001|` in `old` or `new`.
- Prefer the smallest set of edits that solves the task.

The system will apply your JSON edits locally and generate the final unified diff with `git diff`.
Response:

</oracle>
\end{lstlisting}

\subsection{Agent Setting Prompt}
\label{app:prompt-agent}

\begin{lstlisting}[style=promptstyle]
<agent_eval>
You are a software engineer working on a GPU/HPC open-source project.
A task has been assigned to you. Read the description carefully and implement the required changes in the repository.

Focus on:
- Writing correct, working code
- Making minimal, focused changes that address the described requirements
- Building and verifying your changes locally before finishing

Guidelines:
- The workflow has already prepared the base environment before you start
- Reuse the existing environment under `artifacts/` instead of recreating it from scratch
- If you modify compiled code (C++, CUDA, etc.), rebuild and reinstall the package to verify correctness
- You may run the existing test suite at any point to check your work
- Keep all build artifacts, virtual environments, and caches inside the `artifacts/` directory

<task>
{{task}}
</task>
</agent_eval>
\end{lstlisting}

\clearpage
\newpage
\section*{NeurIPS Paper Checklist}

\begin{enumerate}

\item {\bf Claims}
    \item[] Question: Do the main claims made in the abstract and introduction accurately reflect the paper's contributions and scope?
    \item[] Answer: \answerYes{}
    \item[] Justification: The abstract and introduction state the benchmark scope, size, construction, and evaluation settings, and these claims match Sections~\ref{sec:benchmark}--\ref{sec:results}. Comparisons between the oracle and agent settings are stated as empirical measurements of our evaluation configurations.

\item {\bf Limitations}
    \item[] Question: Does the paper discuss the limitations of the work performed by the authors?
    \item[] Answer: \answerYes{}
    \item[] Justification: Section~\ref{sec:discussion} lists the main limitations (a prepared environment, validation on A100/H100 only, the small Perf-Replay subset, LLM-written problem statements, and single pass@1 runs), and Appendix~\ref{app:limitations} discusses further ones.

\item {\bf Theory assumptions and proofs}
    \item[] Question: For each theoretical result, does the paper provide the full set of assumptions and a complete (and correct) proof?
    \item[] Answer: \answerNA{}
    \item[] Justification: The paper has no theoretical results; the only notation defines a task (Section~\ref{sec:formulation}).

    \item {\bf Experimental result reproducibility}
    \item[] Question: Does the paper fully disclose all the information needed to reproduce the main experimental results of the paper to the extent that it affects the main claims and/or conclusions of the paper (regardless of whether the code and data are provided or not)?
    \item[] Answer: \answerYes{}
    \item[] Justification: Sections~\ref{sec:formulation} and~\ref{sec:setup} define the task, the metrics, and the evaluation settings. Appendices~\ref{app:benchmark-details} and~\ref{app:experimental-details} give construction details, model versions, and costs, and Appendix~\ref{app:prompts} gives the prompts. The data and code are public (footnote~2).

\item {\bf Open access to data and code}
    \item[] Question: Does the paper provide open access to the data and code, with sufficient instructions to faithfully reproduce the main experimental results, as described in supplemental material?
    \item[] Answer: \answerYes{}
    \item[] Justification: The benchmark is released at \url{https://huggingface.co/datasets/Froze/SWE-GPU-Bench} and the construction and evaluation code at \url{https://github.com/chenfengshijie/SWE-GPU-Bench}, with instructions for running the evaluation.

\item {\bf Experimental setting/details}
    \item[] Question: Does the paper specify all the training and test details (e.g., data splits, hyperparameters, how they were chosen, type of optimizer) necessary to understand the results?
    \item[] Answer: \answerYes{}
    \item[] Justification: Section~\ref{sec:setup} specifies the oracle and agent settings and the evaluation protocol, and Appendix~\ref{app:models} lists the evaluated models and harnesses. The paper involves no model training.

\item {\bf Experiment statistical significance}
    \item[] Question: Does the paper report error bars suitably and correctly defined or other appropriate information about the statistical significance of the experiments?
    \item[] Answer: \answerNo{}
    \item[] Justification: We report pass@1 rates from a single run per configuration, as in SWE-bench and SWE-Perf, because repeated agent runs on GPU tasks are expensive (Appendix~\ref{app:cost}). We do not report error bars, and Section~\ref{sec:discussion} cautions against reading small differences between models.

\item {\bf Experiments compute resources}
    \item[] Question: For each experiment, does the paper provide sufficient information on the computer resources (type of compute workers, memory, time of execution) needed to reproduce the experiments?
    \item[] Answer: \answerYes{}
    \item[] Justification: Section~\ref{sec:setup} and Appendix~\ref{app:cost} report the GPU type (one A100 or H100 per task) and the token usage and wall-clock runtime of command synthesis and model evaluation.
    
\item {\bf Code of ethics}
    \item[] Question: Does the research conducted in the paper conform, in every respect, with the NeurIPS Code of Ethics \url{https://neurips.cc/public/EthicsGuidelines}?
    \item[] Answer: \answerYes{}
    \item[] Justification: The authors have reviewed the NeurIPS Code of Ethics. The research uses publicly available software-engineering artifacts, focuses on benchmark construction and model evaluation, and includes leakage checks, license safeguards, and isolated evaluation procedures for responsible release.

\item {\bf Broader impacts}
    \item[] Question: Does the paper discuss both potential positive societal impacts and negative societal impacts of the work performed?
    \item[] Answer: \answerYes{}
    \item[] Justification: Appendix~\ref{app:limitations} discusses the positive impact (better agents for maintaining GPU software) and the risk that agent patches introduce numerical or performance regressions, which is why such patches still need review and testing.
    
\item {\bf Safeguards}
    \item[] Question: Does the paper describe safeguards that have been put in place for responsible release of data or models that have a high risk for misuse (e.g., pre-trained language models, image generators, or scraped datasets)?
    \item[] Answer: \answerNA{}
    \item[] Justification: The paper does not release a high-risk model or a sensitive scraped dataset. The benchmark uses only public open-source repositories, and model patches are evaluated in isolated containers (Appendix~\ref{app:limitations}).

\item {\bf Licenses for existing assets}
    \item[] Question: Are the creators or original owners of assets (e.g., code, data, models), used in the paper, properly credited and are the license and terms of use explicitly mentioned and properly respected?
    \item[] Answer: \answerYes{}
    \item[] Justification: All source repositories are public open-source projects, and the release credits each one by recording the source repository of every task.

\item {\bf New assets}
    \item[] Question: Are new assets introduced in the paper well documented and is the documentation provided alongside the assets?
    \item[] Answer: \answerYes{}
    \item[] Justification: SWE-GPU-Bench is documented in Section~\ref{sec:benchmark} and Appendix~\ref{app:benchmark-details} (task format, construction, statistics, and manual review) and is released with a dataset card at \url{https://huggingface.co/datasets/Froze/SWE-GPU-Bench}.

\item {\bf Crowdsourcing and research with human subjects}
    \item[] Question: For crowdsourcing experiments and research with human subjects, does the paper include the full text of instructions given to participants and screenshots, if applicable, as well as details about compensation (if any)? 
    \item[] Answer: \answerNA{}
    \item[] Justification: The work does not involve crowdsourcing experiments or human-subject studies; the human validation described in the appendix is an author/internal expert audit of benchmark artifacts. Therefore participant instructions, screenshots, and compensation details are not applicable.

\item {\bf Institutional review board (IRB) approvals or equivalent for research with human subjects}
    \item[] Question: Does the paper describe potential risks incurred by study participants, whether such risks were disclosed to the subjects, and whether Institutional Review Board (IRB) approvals (or an equivalent approval/review based on the requirements of your country or institution) were obtained?
    \item[] Answer: \answerNA{}
    \item[] Justification: The paper does not involve human-subject experiments or participant data collection; the described human validation is an author/internal expert review of software benchmark artifacts. IRB or equivalent review is therefore not applicable.

\item {\bf Declaration of LLM usage}
    \item[] Question: Does the paper describe the usage of LLMs if it is an important, original, or non-standard component of the core methods in this research? Note that if the LLM is used only for writing, editing, or formatting purposes and does \emph{not} impact the core methodology, scientific rigor, or originality of the research, declaration is not required.
    \item[] Answer: \answerYes{}
    \item[] Justification: LLMs are part of the method: an LLM judge filters pull requests, an LLM writes problem statements, a coding agent synthesizes commands, and the evaluated systems are LLMs (Sections~\ref{sec:construction} and~\ref{sec:setup}, Appendix~\ref{app:prompts}). LLMs were also used to edit the text of the paper and to help visualize results; the authors checked all content.

\end{enumerate}

\end{document}